%% file: main.tex
\documentclass[acmsmall,screen]{acmart}

\AtBeginDocument{%
  \providecommand\BibTeX{{%
    Bib\TeX}}}

\setcopyright{acmlicensed}
\copyrightyear{2018}
\acmYear{2018}
\acmDOI{XXXXXXX.XXXXXXX}

\acmConference[Conference acronym 'XX]{Make sure to enter the correct
  conference title from your rights confirmation emai}{June 03--05,
  2018}{Woodstock, NY}
\acmISBN{978-1-4503-XXXX-X/18/06}

\input{preamble}

\begin{document}



\title{Can We Make Code Green? Understanding Trade-Offs in LLMs vs. Human Code Optimizations}

\author{Pooja Rani}
\orcid{0000-0001-5127-4042}
\email{rani@ifi.uzh.ch}
\affiliation{%
  \institution{University of Zurich}
  \city{Zurich}
  \country{Switzerland}
}

\author{Jan-Andrea Bard}
 \email{jan-andrea.bard@unibe.ch}
\affiliation{%
\institution{University of Bern}
\city{ Bern}
\country{Switzerland}
}

 \author{June Sallou}
 \orcid{0000-0003-2230-9351}
 \email{june.sallou@wur.nl}
\affiliation{%
 \institution{Wageningen University \& Research }
  \city{Wageningen}
\country{The Netherlands}
}

 \author{Alexander Boll}
 \orcid{0000-0002-9881-9748}
 \email{alexander.boll@unibe.ch}
\affiliation{%
\institution{University of Bern}
\city{ Bern}
\country{Switzerland}
}

\author{Timo Kehrer}
 \orcid{0000-0002-2582-5557}
 \email{timo.kehrer@unibe.ch}
\affiliation{%
 \institution{University of Bern}
 \city{Bern, Switzerland}
\country{Switzerland}
 }
 
\author{Alberto Bacchelli}
\orcid{0000-0003-0193-6823}
\email{bacchelli@ifi.uzh.ch}
\affiliation{%
  \institution{University of Zurich}
  \city{Zurich}
  \country{Switzerland}
}
\renewcommand{\shortauthors}{Rani et al.}

\begin{abstract}
The rapid technological evolution has accelerated software development across domains, contributing to a growing share of global carbon emissions.
While large language models (LLMs) claim to assist developers in optimizing code for performance and energy efficiency, 
their real-world efficacy remains 
underexplored--particularly in scientific and engineering domains, where coding practices and environments vary, and green coding awareness is low.
Furthermore, little is known about the optimization strategies they apply or how closely these align with human reasoning.

To address these gaps, we evaluate the effectiveness of LLMs in reducing the environmental footprint of real-world Matlab projects -- a language widely used in both academia and industry for scientific and engineering applications. Unlike prior works on toy programs or benchmarks, we analyze energy-focused optimization on 400 Matlab scripts from 100 top GitHub repositories.
The leading LLMs-- \gpt-3, \gpt-4, Llama, and Mixtral-- alongside a senior Matlab developer, optimized the scripts, which we evaluated on energy consumption, memory usage, execution time consumption, and code correctness.
The developer serves as a real-world baseline for comparing typical human and LLM-generated optimizations.


Mapping 2\,176 proposed optimizations to 13 high-level themes, we find that LLMs propose a broad spectrum of improvements--beyond energy efficiency--including improving code readability \& maintainability, vectorization, memory management, error handling, parallel processing, while the developer overlooked some \eg parallel processing, error handling.
Our statistical tests reveal that 
the energy-focused optimizations unexpectedly had a significant negative impact on memory usage, with no clear benefits in terms of energy consumption or execution time.
We find a strong correlation between execution time and energy consumption across all LLMs, though some outlier cases reveal the trade-offs. 
Our qualitative analysis of these trade-offs indicate that themes such as vectorization, preallocation frequently contributed in shaping these trade-offs.

With LLMs becoming ubiquitous in modern software development, our study serves as a call to action: the need for empirically grounded green coding guidelines.
Additionally, raising energy awareness among developers and LLMs (via training and fine-tuning) is crucial. By bridging these gaps, researchers and developers can better align SE practices with the vision of sustainable software engineering  for real-world applications.

\end{abstract}



\maketitle

\input{content}


\bibliographystyle{ACM-Reference-Format}
\bibliography{references}
\end{document}

%% file: preamble.tex
 \usepackage{amsmath,amsfonts}
 \usepackage{textcomp}
\usepackage{xspace}
\usepackage{float} 
\usepackage[normalem]{ulem} 
\usepackage{url} 

\usepackage{hyperref}
\usepackage{subcaption}
\usepackage{tablefootnote}
\usepackage{array, booktabs}
\usepackage{blindtext}
\usepackage{ifthen}
\usepackage[shortlabels]{enumitem}

\usepackage{siunitx}

\usepackage{csvsimple}
\usepackage{supertabular} 
\usepackage[most]{tcolorbox}

\usepackage{multirow}
\usepackage{tabularx}
\usepackage{lscape} 

\usepackage{fontawesome}      
\usepackage{adjustbox} 		
\usepackage{pifont} 		

\newcolumntype{R}[2]{
    >{\adjustbox{angle=#1,lap=\width-(#2),margin*=0.4em 0em 0em 0em}\bgroup}
    l
    <{\egroup}
}

\let\oldFootnote\footnote
\newcommand\nextToken\relax
\renewcommand\footnote[1]{%
    \oldFootnote{#1}\futurelet\nextToken\isFootnote}
\newcommand\isFootnote{%
    \ifx\footnote\nextToken\textsuperscript{,}\fi}

\newcommand{\checkdecimal}[1]{%
    \IfDecimal{#1}{#1}{0}%
}

\definecolor{gray50}{gray}{.5}
\definecolor{gray40}{gray}{.6}
\definecolor{gray30}{gray}{.7}
\definecolor{gray20}{gray}{.8}
\definecolor{gray10}{gray}{.9}
\definecolor{gray05}{gray}{.95}

\newlength\Linewidth
\def\findlength{\setlength\Linewidth\linewidth
	\addtolength\Linewidth{-4\fboxrule}
	\addtolength\Linewidth{-3\fboxsep}
}
\newenvironment{rqbox}{\par\begingroup
	\setlength{\fboxsep}{5pt}\findlength
	\setbox0=\vbox\bgroup\noindent
	\hsize=0.95\linewidth
	\begin{minipage}{0.95\linewidth}\normalsize}
	{\end{minipage}\egroup
	\textcolor{gray20}{\fboxsep1.5pt\fbox
		{\fboxsep5pt\colorbox{gray05}{\normalcolor\box0}}}
	\endgroup\par\noindent
	\normalcolor\ignorespacesafterend}

\newcommand{\rb}[1]{
	\begin{tcolorbox}[colback=gray!05,
		colframe=black,
		width=\columnwidth,
		arc=3mm, auto outer arc,
		boxrule=0.5pt,
		]
		#1
	\end{tcolorbox}
}

\newcounter{Finding}
\stepcounter{Finding}

\newcommand{\roundedbox}[1]{
	\rb{
		\noindent
		{\textbf{Finding \theFinding}. #1}
	}
	\stepcounter{Finding} 
}

\newcommand\pr[1]{\nbc{PR}{#1}{violet}}

\newcommand{\ie}{\emph{i.e.},\xspace}
\newcommand{\eg}{\emph{e.g.},\xspace}
\newcommand{\etal}{\emph{et al.}\xspace}
\newcommand{\etc}{\emph{etc.}\xspace}
\newcommand{\github}{{GitHub}\xspace}

\newcommand{\gpt}{GPT\xspace}
\newcommand{\llama}{Llama\xspace}

\newcommand{\rqIIheader}{Energy Optimization Themes\xspace}
\newcommand{\rqIIIheader}{Impact of Energy Optimizations by LLM \& Human \xspace}

\newcolumntype{L}[1]{>{\raggedright\arraybackslash}m{#1}} 

\def\BibTeX{{\rm B\kern-.05em{\sc i\kern-.025em b}\kern-.08em
    T\kern-.1667em\lower.7ex\hbox{E}\kern-.125emX}}
    
\newboolean{showedits}
\setboolean{showedits}{true} 
\ifthenelse{\boolean{showedits}}
{
	\newcommand{\del}[1]{\textcolor{red}{\sout{#1}}} 
	\newcommand{\nbe}[3]{
		{\colorbox{#3}{\bfseries\sffamily\scriptsize\textcolor{white}{#1}}}
		{\textcolor{#3}{\sf\small$\blacktriangleright$\textit{#2}$\blacktriangleleft$}}}
}{
	\newcommand{\del}[1]{} 
	
	\newcommand{\nbe}[3]{}
}

\ifthenelse{\boolean{showedits}}
{
	\newtcolorbox{inserted}{
		title=Inserted text:,
		colframe=blue,colback=blue!5!white,
		breakable,
		leftrule=0mm, 
		bottomrule=0mm,
		rightrule=0mm,
		toprule=0mm,
		arc=0mm, outer arc=0mm,
		oversize
	}
	\newtcolorbox{deleted}{
		title=Deleted text:,
		colframe=red,colback=red!5!white,
		breakable,
		leftrule=0mm, 
		bottomrule=0mm,
		rightrule=0mm,
		toprule=0mm,
		arc=0mm, outer arc=0mm,
		oversize
	}
	\newtcolorbox{refactored}{
		title=Rewritten text:,
		colframe=blue,colback=red!5!white,
		breakable,
		leftrule=0mm, 
		bottomrule=0mm,
		rightrule=0mm,
		toprule=0mm,
		arc=0mm, outer arc=0mm,
		oversize
	}
}{

}

\newboolean{showcomments}
\setboolean{showcomments}{true}
\ifthenelse{\boolean{showcomments}}
{\newcommand{\nbc}[3]{
		{\colorbox{#3}{\bfseries\sffamily\scriptsize\textcolor{white}{#1}}}
		{\textcolor{#3}{\sf\small$\blacktriangleright$\textit{#2}$\blacktriangleleft$}}}
	}
{\newcommand{\nbc}[3]{}
	}

\definecolor{source}{gray}{0.9}
\newboolean{isblinded}
\setboolean{isblinded}{true}
\ifthenelse{\boolean{isblinded}}
{\newcommand\blind[1]{BLINDED\xspace}}
{\newcommand\blind[1]{#1\xspace}}

\newcommand{\rqI}{What are the challenges in establishing an automated pipeline for measuring the energy consumption in Matlab projects and how can they be addressed?}
\newcommand{\rqII}{What guidelines do LLMs and developer recommend for energy optimizations?}
\newcommand{\rqIII}{Do the recommended optimizations impact code's energy efficiency?}


%% file: content.tex
\section{Introduction}
The rapid evolution of technology, digital transformation, and data-intensive analytics has spurred software development across various domains, contributing significantly to carbon emissions worldwide \cite{CleanEnergy,jumper2021highly}.
It is projected that the global information \& communication technology (ICT) electricity consumption could reach up to 30,715TWh (51\% of total consumption) by 2030~\cite{andrae2015global}.
While the software engineering (SE) research community has started investigating energy efficiency for various domains, \eg mobile and web applications~\cite{manotas2016empirical,rani2024energy}, and programming languages~\cite{pereira2017energy}, many other domains and systems (\eg simulation applications) remain largely unexplored. This includes systems built by domain experts 
such as scientists and engineers \cite{silvia2017user},
who have limited awareness of SE best practices (\eg maintainability and usability~\cite{carver2007software,gonzalez2020state}).
They often lack access to guidelines and tools for optimizing energy consumption in their code, have different needs, and follow diverse coding practices that further complicate optimization efforts~\cite{abdulsalam2014program,pang2015programmers,manotas2016empirical,schneider2022transformation2,rani2024energy}.

Recent advances in large language models (LLMs) claim to assist developers in tasks, such as refactoring code, fixing bugs, and optimizing code~\cite{chen2024supersonic,sadik2023analysis,hou2023large,florath2023llm,vartziotis2024learn}, making them more appealing to domain experts (or non-software engineers) seeking efficient code generation or code refactoring.
Given that even small code optimizations can influence energy efficiency~\cite{pathak2012energy,hasan2016energy,singh2015impact}, 
 evaluating LLMs' ability to promote sustainable or energy-efficient code in real-world scenarios is critical.
However, to our knowledge, no research has assessed the  effectiveness of LLMs in generating energy-efficient code in such domains~\cite{vartziotis2024learn,sanchez2024controlled,cappendijk2024generating}. 
Moreover, little is known about the types of optimizations LLMs propose—or how closely they align with those applied by human developers in practice.
Existing research has merely scratched the surface, revealing LLMs' limited sustainability awareness ~\cite{vartziotis2024learn} and their focus on trivial coding tasks or specific languages~\cite{cursaru2024controlled,cappendijk2024generating}.

To address these gaps, we evaluate the energy-efficiency potential of four LLMs (both open-source and closed-source LLMs)-- GPT-3, GPT-4, Llama, Mixtral across 400 scripts from 100 real-world GitHub projects. 
We focus on Matlab as a representative language, widely used  by millions of engineers and scientists in academia and industry~\cite{higham2016matlab,leroy2022role,wilson2014best,tomaszewski2023analyzing,silvia2017user,matlabPendulum2024}, 
making it a relevant case study within this broader research gap.
It is a multi-paradigm programming language and multi-purpose platform, developed by MathWorks\footnote{\url{https://www.mathworks.com/}} for complex simulations and modeling applications, AI-driven and data-intensive systems, which often are computationally intensive and energy consuming~\cite{leroy2022role,matlabPendulum2024,matlabBioinfo2024,wilson2014best,tomaszewski2023analyzing}. 
We also involved a senior Matlab developer, having general expertise in Matlab but not in energy optimization, serves as a realistic baseline for current human practices, allowing us to compare LLM-generated optimizations with those of a typical experienced developer.
We evaluate the optimized scripts on the metrics such as energy consumption, execution time, memory usage, and code correctness.
%
We assess the impact of these optimizations using statistical tests
and conduct a qualitative analysis of recommendations from LLMs and the developer to identify various energy-related themes or patterns~\cite{rani2024energy}.
Furthermore, we analyze a sample of optimized scripts to identify the potential themes influencing energy or execution time efficiency.

Each LLM optimized 400 Matlab scripts, identifying 2,176 unique optimizations mapped to 13 high-level themes. 
The developer manually optimized 53 scripts based on his experience and established guidelines. 
Although instructed for energy efficiency, LLMs addressed a broad range of themes -- from improving the readability \& maintainability of code to memory management to error handling.
Similarly, the developer focused on some similar themes (not to the same extent) and neglected others like parallel processing, error handling, plots, or comments.
Unexpectedly, we found that these optimizations do not significantly improve energy, memory, or execution time.
Statistical tests showed a significant negative impact on memory usage without any apparent energy consumption or execution time benefits.
Specifically, \gpt-4 optimizations increased energy, memory, and time consumption, indicating a lack of energy efficiency awareness while \llama and Mixtral showed a positive effect on energy consumption.
Across all models, we observed a overall strong correlation between energy consumption and execution time, consistent with prior work \cite{pereira2017energy,pinto2014understanding,roque2024unveiling,strubell2020energy}. However, several outlier cases revealed trade-offs, where gains in one metric came at the cost of another. To explore this trade-offs, we qualitatively analyzed scripts in the top and bottom quartile of optimization differences (efficient and inefficient ones).
We identified recurring themes including vectorization, preallocation, function replacement, and code reorganization that contributed to efficiency gains in one dimension while degrading another.
Our qualitative analysis of energy-intensive scripts revealed energy anti-patterns that may have contributed to it, such as inefficient calculation and redundant function evaluation.

Overall, we conclude that LLMs currently provide minimal support for improving energy efficiency of Matlab code, aligning with findings in other languages~\cite{vartziotis2024learn,cappendijk2024generating,sanchez2024controlled}.
Our qualitative analysis show that they aim to enhance code quality (\eg readability, maintainability).However, their impact on energy efficiency remains ineffective, with frequent trade-offs among energy, execution time, memory, and code correctness.
Future work should investigate diverse coding practices to offer evidence-based recommendations to developers for energy-efficient coding.
With this study, we make the following main contributions:
\begin{itemize}
    \item Our automated pipeline to measure Matlab code's efficiency~\cite{reppackage}
    \item A list of energy optimization themes proposed by LLMs and a senior developer
    \item A quantitative and qualitative analysis of optimizations and their impact on various efficiency measures
    \item Our dataset of curated Matlab projects available in the replication package (RP)~\cite{reppackage}
\end{itemize}

\section{Research Questions}
This section motivates our research questions and outlines the method to answer them.

With software increasingly contributing to carbon emission \cite{CleanEnergy,jumper2021highly}, optimizing energy efficiency has become crucial \cite{dutta2023case}, especially in energy-intensive domains such as AI and data-driven systems \cite{wilson2014best,tomaszewski2023analyzing,niazai2023applications}.
Many of these systems are developed by non-software engineers with limited awareness of SE best practices~\cite{silvia2017user,carver2007software}.
LLMs claims to support developers in various SE tasks, including refactoring code, or optimizing code for energy consumption or execution time ~\cite{chen2024supersonic,sadik2023analysis, hou2023large,vartziotis2024learn,sanchez2024controlled,cappendijk2024generating}, making them more appealing to domain experts (non-software engineers) in generating energy efficient code.
However, little is known about the optimization guidelines LLMs apply, and how closely these align with human optimization.
Therefore, we asked LLMs and the developer to optimize a sample of Matlab scripts from the top 100 real-world \github projects and to provide the guidelines. We mapped the guidelines to high-level themes and compared the themes from LLMs and a senior developer, asking:
\begin{center}
	\begin{rqbox}
		\begin{description}
			\item $RQ_1$: \rqII
		\end{description}
	\end{rqbox}
\end{center}

Previous studies show mixed results in leveraging LLMs for  code efficiency but often focused on isolated aspects, such as execution time, energy, resource usage or task accuracy~\cite{wright2023efficiency,Luis19a}.
While Florath finds LLMs effective in optimizing execution time efficiency for Python code~\cite{florath2023llm}, others highlight their limitations in generating energy-efficient code~\cite{vartziotis2024learn,cappendijk2024generating,cursaru2024controlled}.
Moreover, they primarily evaluated LLMs on small-scale coding problems from interview platforms rather than code from real-world projects~\cite{cursaru2024controlled,cappendijk2024generating}.
Therefore, we evaluate LLM- and developer-suggested optimizations on the real-world Matlab scripts, assessing their impact in terms of energy consumption, execution time, memory usage, or code correctness. Additionally, we analyze the potential themes (from RQ1) that might have contributed to energy inefficiency and compare LLM and developer optimizations against the original code.
\begin{center}
	\begin{rqbox}
		\begin{description}
			\item $RQ_2$: \rqIII
		\end{description}
	\end{rqbox}
\end{center}

\section{Methodology}
\label{sec:methodology}
\begin{figure*}[tbh]
    \centering
    \includegraphics[width=0.95\linewidth]{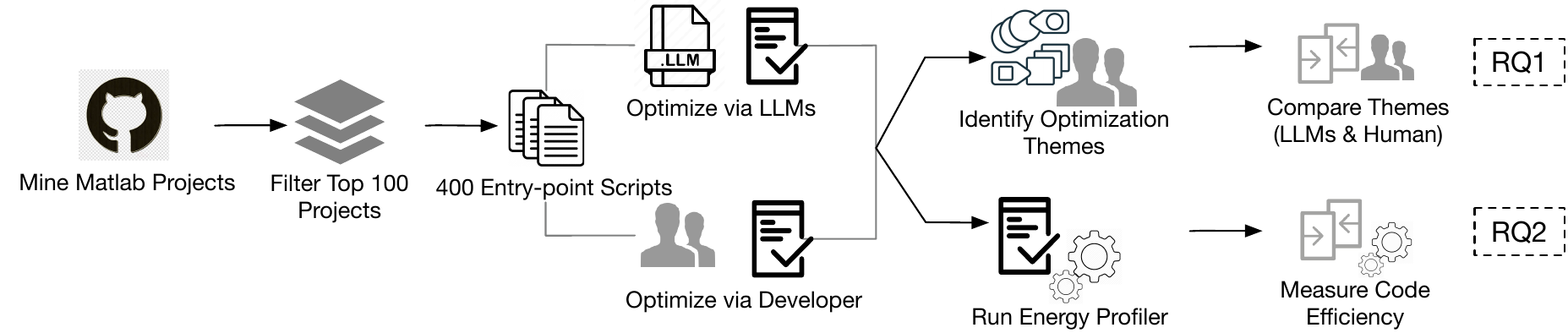}
    \caption{Study design and overall steps followed to answer all research questions.}
    \label{fig:study-design}
\end{figure*}

\textit{Language selection:}  Despite growing attention to green coding practices~\cite{manotas2016empirical,rani2024energy}, many domains remain largely unexplored.
Matlab is a popular platform widely used by millions of engineers and scientists in academic and industrial projects in various domains~\cite{higham2016matlab,leroy2022role,wilson2014best,tomaszewski2023analyzing,silvia2017user,matlabPendulum2024}.
It is developed by MathWorks for developing computational mathematics-based applications, models, or AI-driven systems~\cite{higham2016matlab,leroy2022role,wilson2014best,tomaszewski2023analyzing,silvia2017user,matlabPendulum2024}, which can be computationally expensive.
While the scientific community has prioritized hardware energy optimization over software-level optimization\cite{lacoste2019quantifying,lannelongue2021green,jumper2021highly,grealey2022carbon}, the SE community focused on mobile and web applications written in general purpose languages like Python, Java, or Kotlin. To our knowledge, our study is the first to examine energy efficiency of Matlab code.
%
\autoref{fig:study-design} shows the overall steps followed to answer our research questions.

\textit{Project selection:} 
Previous studies in the MathWorks ecosystem have curated datasets, \eg SLNET and EVoSL~\cite{shrestha2022slnet,10343917} but they did not focus on Matlab projects.
Therefore, we mined our own dataset of open-source projects from \github using the following inclusion and exclusion criteria.\\
\textit{Inclusion criteria:}
\setlist{nolistsep}
\begin{enumerate}[label=I\arabic*:, start=1, itemsep=0em]
\item
Matlab is the primary language of the project (using the \textit{Linguist} API of \github
~\cite{githublinguist2024});
\item The project is dated between Jan 2008 and Feb 2024 to ensure the scripts of Matlab are runnable with the Matlab version R2021b;\footnote{This version was chosen to match the organization's Matlab license for running experiments on remote servers.}
\item The project has at least two contributors to ensure that it is not a personal toy program, and 
\item The project contains a runnable Matlab script -- otherwise we cannot measure any energy consumption for this project.
\end{enumerate}
\textit{Exclusion criteria:}
\begin{enumerate}[label=E\arabic*:, start=1, itemsep=0em]
\item The project is not maintained anymore, and
\item The project is a mirror repository.
\end{enumerate}

Based on the inclusion and exclusion criteria, we identified 372,388 \github projects.
After filtering for inactive projects and excluding projects by the MathWorks company itself, 371,994 remained.
Then, we narrowed down this list to projects of the highest quality: we considered the top 10\%\footnote{If a higher quantile of projects scored the same as the 10th percentile on a metric, we excluded it as well.}
based on contributors (68,070), stars (24,255), commits (20,857), length of project lifetime (18,771), and project activity (15,203), narrowing the list to 15,203 high quality projects~\cite{reppackage}.
Since our aim is to get a significant sample of scripts to establish our automated pipeline, we selected and cloned the top 100 popular repositories based on their number of stars.

\textit{Entry-point identification:} 
After cloning the projects, we aimed to identify the main script(s) or entry-point(s) of the projects to eventually run them automatically for measuring their energy usage.
It is often challenging to identify which scripts users would execute at first.
We reviewed the project documentations, specifically README files.
When documentation was missing, we established various heuristics to search for executable \emph{entry-point} scripts automatically:
 entry-point scripts may call other scripts but never be called by other scripts.
We generated a directed forest for the calling relationship among all Matlab scripts of a project. In this forest, edges represent one script calling another, and nodes represent the scripts~\cite{reppackage}. 
The roots of the forest were identified as entry-point scripts.
Furthermore, we filtered out entry-points that required input parameters, as guessing suitable input values is a tedious to impossible task. 
After this filtering step, 1,937 entry-point scripts remained for further analysis.
After testing their execution without crashes, we obtained a set of 400 entry-point scripts, representing 47 out of the 100 projects~\cite{reppackage}.

 \textit{Energy profiler:} 
To measure the performance and energy consumption of code, we explored various tools.
Although, MathWorks provides a profiler to measure the performance of source code~\cite{matlab2024}
it lacks energy measurement capabilities. 
Jay \etal \cite{jay2023experimental} compared several such power measurement tools that can handle CPU or GPU-based infrastructure. For example, \emph{PowerAPI} provides per-process energy consumption, while \emph{RAPL}, an Intel-technology, measures per-component (CPU, DRAM) energy consumption. 
We utilized the \emph{RAPL}-based tool \emph{EnergiBridge} \cite{sallou2023energibridge} as, it is a platform-independent tool and has been already used for various previous works~\cite{roque2024unveiling}. 
 It provides a unified interface for energy measurements, allowing our pipeline to run across various platforms and operating systems. Using EnergiBridge, we built an automated pipeline to measure the energy consumption of the different versions of Matlab scripts.


\subsection{\texorpdfstring{$RQ_1$}{RQ1}: \rqIIheader}
\label{subsec:rqII-methdology}
This RQ focuses on analyzing guidelines for improving the energy efficiency of 400 entry-point scripts using LLMs and a senior Matlab developer. We refer to these 400 scripts as \emph{`original scripts'} and their corresponding modified (optimized) versions as the \emph{`optimized scripts'}.
For the LLM optimization, we instructed the models to provide the reasoning for the applied optimization. 
We manually analyzed the frequently appearing unique themes of optimization suggestions and categorized them into high-level themes using thematic analysis and closed-card sorting~\cite{braun2006using,coleman2007using}.
For the developer optimization, we first established energy optimization guidelines  with the Matlab developer (with ten years of experience). 
Then, the developer applied those guidelines to randomly sampled 53 scripts (out of 400).


    
 \textit{Selected LLMs:}
To assess various energy optimization strategies of recent models, we utilized popular, widely-used LLMs, both open-source and closed-source~\cite{ateia2024can}.
For the closed-source models, we used \gpt-3.5-turbo and \gpt-4o~\cite{radford2018improving,achiam2023gpt} due to their great performance in various tasks~\cite{florath2023llm}, and widespread usage in industry and academia ~\cite{ateia2024can,hou2024benchmarking}.
To assess \gpt models, we used the API provided by OpenAI to generate optimized code based on the prompt. 
    We configured the model's three parameters: \textit{max\_tokens}, \textit{temperature}, and \textit{top\_p}.
    We set the \textit{max\_tokens} to the maximum value 4096 or 8192 respectively as supported by the models, \textit{temperature} to 0.5 as its lower values make the model’s responses more predictable and deterministic~\cite{APIParameters}.
    We set \textit{top\_p}, also known as nucleus sampling, to 1 as it determines the number of possible words to consider. Higher values indicate a richer vocabulary.
     For the open-source models, we used the \llama model (\ie Meta-Llama-3-70B-Instruct) from Meta and the Mixtral model (\ie Mixtral-8x22B-Instruct-v0.1)~\cite{jiang2024mixtral} from Mistral AI. These open-source models have shown competitive performance with \gpt-based models~\cite{ateia2024can}, outperforming them on several benchmarks~\cite{ateia2024can,hou2024benchmarking}. We used the API endpoint from Anyscale
     to access these models~\cite{anyscale2024} with configuration settings similar to the ones used by the GPT models. \footnote{We accessed the open-source models and closed-source models via Anyscale in June 2024.} 

\begin{table}[h!]
    \centering
    \begin{minipage}{0.58\textwidth}
        \input{Listings/Prompt-llm.tex}
    \end{minipage}
    \hfill
    \begin{minipage}{0.4\textwidth}
        \input{Listings/optimized-code-GPT-3.5}
    \end{minipage}
\end{table}

\textit{Prompt engineering:}
Previous work has explored various strategies to modify prompts to generate energy-efficient code~\cite{florath2023llm,cappendijk2024generating,vartziotis2024learn}. 
Following their strategies, our prompt also focused on stating the goal directly, \emph{``Optimize the code for energy efficiency''} to ensure that the LLM response is aligned with our goal of generating energy-efficient code.
We also provided a system prompt to ensure that the model acts as an expert.
We instructed the model to provide the reasoning for the applied code changes and thus collected energy optimization guidelines.
We refined the prompt to output the Matlab code in a specific format for easier processing.
For instance, \autoref{lst:Prompt-llm}
is provided to the \gpt-3 model, and its response is shown in \autoref{lst:optiimized-PSO-function-gpt3}.

\textit{Guideline categorization:}
Next, we categorized the reasoning of the LLMs (\eg shown in \autoref{lst:optiimized-PSO-function-gpt3}) in a semi-automated way (keyword based and manual analysis) to better understand the types of themes LLMs apply for energy optimization.
Since multiple optimizations can be applied to each file, several themes may be present per file. 
First, we automatically extracted line headings from each script (if there was any) for each optimization. 
For example, from the \gpt-3 response in \autoref{lst:optiimized-PSO-function-gpt3}, we extracted themes as: \emph{parameterization}, and \emph{loop optimization}.
Despite being instructed in the prompt, we observed that the models provided the reasoning in varying output formats. For instance, Mixtral gave a line of reasoning for every small change made to the original code, while \gpt models grouped similar changes and used a line header for the themes as shown in \autoref{lst:optiimized-PSO-function-gpt3}.
To ensure reliable theme extraction, we manually verified ten randomly selected scripts per model and refined the prompt to format output code.

Then, we first merged similar themes across all scripts, applying a similarity threshold $>$ 0.8 using the NLTK library.\footnote{\url{https://www.nltk.org/}} The themes were then sorted based on how frequently they were mentioned. The first author manually grouped the frequent themes (occurring more than once) into high-level categories using thematic analysis~\cite{braun2006using,coleman2007using}. 
The initial high-level themes are derived from the LLMs' most frequent themes using the closed-card sorting method~\cite{braun2006using,coleman2007using}, then refined and merged~\cite{reppackage}. For example, themes related to simplifying code, improving formatting, and restructuring were grouped under \emph{improved code readability \& maintainability}.
While manually grouping themes, the author noticed certain keywords associated with various themes, \eg \emph{preallocation}, \emph{pre-computation}, or \emph{improved memory management}. Thus, we composed additional rules to automate theme mapping (reported in RP~\cite{reppackage}). 
For each model, we aimed to manually address at least the themes mentioned more than once and account at least half of the total mentions (\eg 500 mentions out of 1000).

\textit{Energy-related guidelines in Matlab:}
The developer first explored whether energy optimization guidelines exist in Matlab and found none.
We reached out to the MathWorks support team regarding  energy optimization guidelines or tools and were redirected to the official guidelines and Matlab profiler recommendations. However, none of these sources also included any energy-related guidelines.
Based on Matlab profiler recommendation and his experience, the developer composed diverse energy optimization guidelines together with their reasoning (reported in RP~\cite{reppackage}).
The first author verified and categorized them into the high-level themes previously formed for the LLMs' reasoning.
Finally, we manually compared these optimization themes from LLMs and the developer. 

\subsection{\texorpdfstring{$RQ_2$}{RQ2}: \rqIIIheader}
\label{subsec:rqIII-methdology}
To answer this RQ, we evaluated the energy efficiency of the original and optimized code by executing the 400 scripts in our energy measurement setup as shown in \autoref{fig:study-design}. 
We assessed the execution results based on various metrics, \eg execution time, energy, memory, task accuracy~\cite{florath2023llm,vartziotis2024learn}. 

\textit{Project execution:} 
To execute the projects automatically, we added all standard Matlab toolboxes\footnote{\url{https://github.com/mathworks-ref-arch/matlab-Dockerfile/blob/main/mpm-input-files/R2021b/mpm_input_r2021b.txt}} to our execution environment, as some projects depended on them.
Since each project can have several dependencies, possibly causing conflicts,
we prepared a Docker image to run each project independently in its own container.
The Docker image is prepared using the official Docker image provided by MathWorks.
Each Docker image is run on our university's server (configuration provided in RP).
Following prior work~\cite{rani2024energy,sallou2023energibridge}, we ran each project 30 times to control for measurement variability. Further measures within our testing protocol include:
\begin{enumerate}[label=P\arabic*:, start=1, noitemsep]
\item A 5-second pause (cool-down period) interposed between each run to ensure hardware temperature consistency and to prevent residual energy from the previous run,
\item The order in which each script is executed and measured has been randomized,
\item The projects are run in a Docker image,
\item It is ensured that non-essential applications are not running and interacting,
\item The sleep mode has been disabled (set the sleep time to never).
\end{enumerate}

\textit{Evaluation:}
Most prior studies on energy consumption focused on processor energy (\texttt{replCPU\_Energy}) \cite{rani2024energy}. 
Barroso \etal emphasized the need to address energy use in DRAM, storage, and networking components\cite{barroso2019datacenter}. For instance, they noted that high RPM disk drives can use up to 70\% of their power just to keep the platters spinning.
Therefore, we measured \emph{memory usage} alongside \texttt{CPU\_Energy}.
Building on the work of Vartziotis \etal \cite{vartziotis2024learn}, we adapted the \emph{green capacity} metric based on various code efficiency metrics of interest:
\begin{itemize}
    \item\textit{Code correctness}: Assess whether the optimized code preserves the original script output.
    \item\textit{CPU energy consumption}: Measure the amount of energy consumed during its execution -- this is reported in Joules (J). Lower values indicate a more energy-efficient code. 
    \item\textit{Execution time}: Measure the time taken by the code execution. It is measured in milliseconds (ms). Lower values indicate a time-efficient code.
    \item\textit{Memory usage}: Measure the DRAM taken by the program during its execution. It is measured in kilobytes (kB). Lower values indicate a more memory-efficient code.
     \item \textit{Performance delta (PD)}: 
     Compute the difference between the original and the optimized code for each metric (energy consumption, memory, or execution time). A positive value indicates an improvement (reduction) in the respective metric, while a negative value indicate a deterioration.
\end{itemize}
For code correctness, we executed each original and optimized script pair-wise, comparing: i) their console output, ii) script changes in the file system coming from script execution, and iii) the return values. In a first step, we flagged non-deterministic scripts, that produced differences for any of these three characteristics in multiple runs. Such non-deterministic scripts were not evaluated for correctness. For deterministic scripts, a script was classified as `correct' if it produced identical results for all three characteristics. 
For the other metrics, we used EnergiBridge to collect energy consumption per core (\texttt{CPU\_Energy}), memory usage 
(\texttt{USED\_memory}), and execution time (\texttt{initial\_time - end\_time}).

\textit{Statistical tests:}
To test our hypothesis that LLM optimizations positively impact code's energy efficiency, we employed statistical hypothesis tests.
We applied a Shapiro-Wilk test to the original metric values to check for normal distribution. We found that the metrics were not normally distributed, therefore, we employed a non-parametric Wilcoxon test (paired test) as each script was independently selected, executed, and tested under both conditions (with and without optimization). 
We then assessed the statistical power to ensure a sufficient likelihood of detecting a true effect. An \emph{a priori} power analysis using G-power \cite{faul2007g} for a two-tailed test with $\alpha=0.05$, $\beta=0.95$ and $dz=0.02$ (aiming to detect medium to large effects for LLMs), indicated a minimum sample size of 343.
From the previous RQ, we obtained a set of 400 scripts.
We optimized and executed these 400 scripts and observed that some were not optimized by LLMs, while others crashed due to timeout or input errors. 
After removing these, we had at least 359 valid scripts per model (\gpt-3: 360, \gpt-4: 359, \llama: 369, Mixtral: 367), all satisfying our minimum sample size criterion of 343 scripts.

For the human-focused optimization, the developer optimized a smaller sample of 53 scripts due to the task being time and effort-intensive (ca. 20 hours in total). We expected the senior developer to make focused code changes, resulting in a bigger impact on code's performance. The developer had no time constraints, allowing for thorough optimization without time pressure~\cite{mantyla2013more}.

To choose the appropriate statistical test for the metrics, 
we compared the median of each metric for the optimized scripts against the original scripts to 
determine if the optimized versions tend to perform equally or better.
Based on the results, we used either a two-sided or one-sided Wilcoxon-signed rank test.
Additionally, to check if these metrics (\eg energy and memory) were correlated, we used the Spearman correlation coefficient \cite{Spearman}.

\textit{Qualitative analysis of themes and trade-offs:} 
To analyze the impact of optimizations, two authors conducted a qualitative study on the LLM- and human-optimized scripts. Since multiple themes were applied on each file, isolating their individual effect was challenging. However, we identified key themes influencing various metrics. 
Specifically, we examined cases where energy and execution time trends diverged in the optimized scripts. Using a diff-based analysis, we manually compared code changes between the original and optimized versions.
For LLM optimizations, we analyzed the distribution of Performance Delta, focusing on the upper (top 25\%) and lower (bottom 25\%) quartiles of the metrics to ensure scripts indeed had differences and to capture strong trends in optimization effects.
For human optimizations, we identified the cases of energy and time metrics moving in the same direction and the opposite direction. 
To further explore specific code changes, we followed Cruz’s guidelines \cite{cruz2021green}, running randomized experiments 30 times with a 5-second pause between tests on a Lenovo ThinkPad P16s Gen 2 (Ubuntu 24.04). 

\textit{Our experiment footprint:} Following the Green SE guidelines~\cite{cruz2021green}, we also estimated our study's footprint.
Our study intends to measure the energy footprint of Matlab code so we estimated the energy usage of our experiment, including the energy consumed by LLM APIs per request. Previous studies have shown that the operating cost of Chat\gpt is 564 MWh per day for 1195 million requests~\cite{desislavov2023trends}.\footnote{\url{https://www.baeldung.com/cs/chatgpt-large-language-models-power-consumption}} This yields an average of 0.472Wh per request. Since data from AnyScale is unavailable, we applied the same estimate for other models like \llama and Mixtral.
With four LLMs used per request, each script optimization consumed 1.888 Wh. 
Adding a 50\% safety margin, the energy per request was estimated at 2.9 Wh (10,440 J), resulting in 1.16 kWh for 400 requests or scripts.
Our server consumption is estimated at 83.8~W at idle time (no specific workload running) and 111.5~W for a workload of 10 inference LLM requests (\ie llama model)~\cite{ThesisGreenServerNumber}. 
Using the estimated average power consumption of  97.65~W,
the server ran about one week, consuming 16.4 ~kWh. 
Repeating the experiment twice resulted in a total estimated consumption of at least 35.06~kWh. To put it into perspective, it is equivalent to around 1.2 day of electricity usage for an average American household.\footnote{\url{https://www.energybot.com/blog/average-energy-consumption.html}}

\section{Result and Discussion}

\subsection{\texorpdfstring{$RQ_1$}{RQ1}: \rqIIheader}
\label{subsec:rqII-results}


In this RQ, we provided 400 Matlab scripts to various LLMs to optimize for energy efficiency, and explain their reasoning as shown in \autoref{lst:Prompt-llm}. 
An example response from \gpt-3 is given in \autoref{lst:optiimized-PSO-function-gpt3}.
\input{Tables/Themes-frequency}

\input{Tables/Optimization-categories}

\begin{figure}
    \centering
    \begin{subfigure}{0.49\linewidth}
        \includegraphics[width=0.99\textwidth]{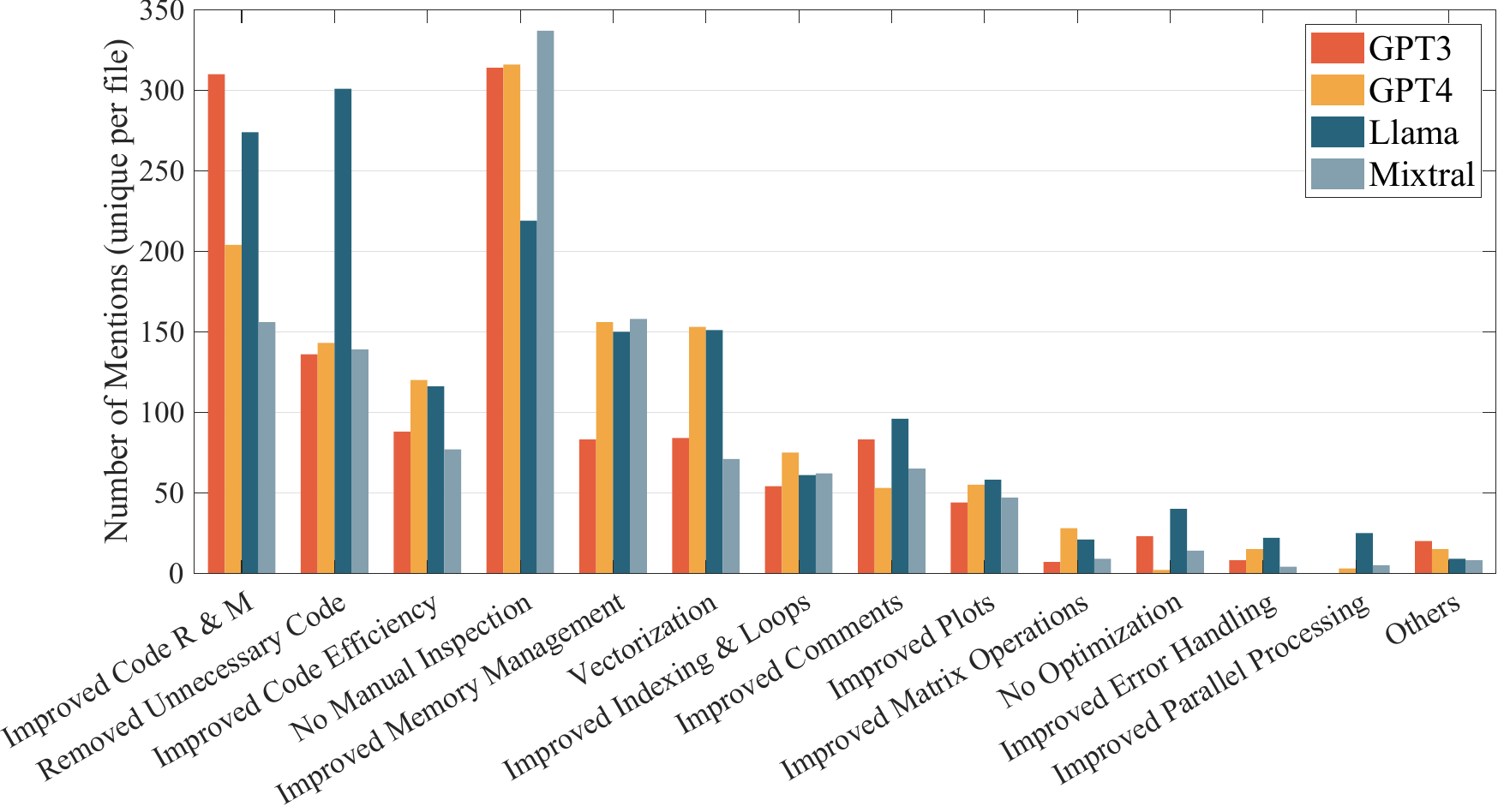}
        \caption{Theme frequency in LLM-optimized scripts}
        \label{fig:themes-all-models}
    \end{subfigure}
    \hfill
    \begin{subfigure}{0.49\linewidth}
        \includegraphics[width=0.99\textwidth]{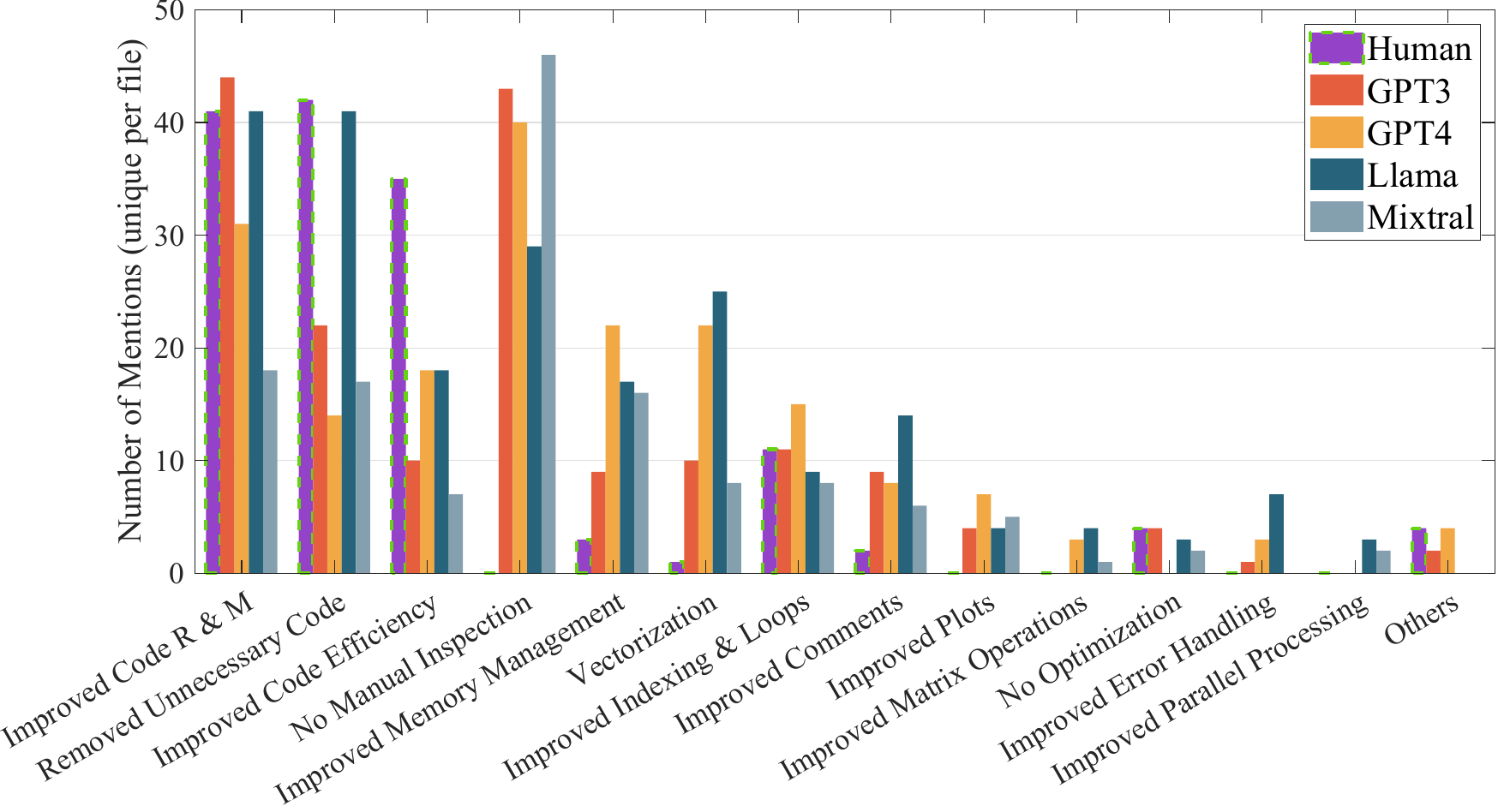}
        \caption{Theme frequency in human-optimized scripts}
        \label{fig:themes-human-optimized-Files}
    \end{subfigure}    
    \caption{Theme ``Improved Code R \& M'' stands for ``Improved Code Readability \& Maintainability''.}
\end{figure}

We first analyzed the reasoning provided by LLMs for energy optimization, identifying 3,843 themes. 
\autoref{tab:reasoning-themes-frequency} shows the unique and total themes extracted from the provided reasoning in the scripts.
Mixtral suggested the most themes (1,335), and \llama the fewest (703).
We manually mapped the frequently appearing low-level themes (2,176 out of 3,843) to 13 high-level categories per model, such as \emph{improved indexing \& loops}, as described in \autoref{tab:optimization-categories} and their frequency is shown in \autoref{fig:themes-all-models}.
We observed that \gpt-3 and \gpt-4 suggested similar optimization themes.
Mixtral provided code-specific themes for every line change in the code, such as, the theme \emph{``changed the way fields are accessed by using the dot notation instead of parentheses''} is mentioned in 133 scripts.
\llama provided general themes such as \emph{removed unnecessary code} for energy optimization. 
The complete list of themes, along with their mapping, are provided in the RP \cite{reppackage}.

\roundedbox{
Although instructed to optimize for energy efficiency, 
LLMs focused on a broad range of maintainability and performance-related themes \eg \emph{improving readability \& maintainability}, \emph{improving code efficiency}, \emph{parallel processing}, or \emph{error handling}. In contrast, the developer often focused more narrowly on maintainability and efficiency.
}

\autoref{fig:themes-all-models} shows the frequency of these themes across all models even though they were instructed to optimize energy efficiency, particularly \emph{improving code readability \& maintainability} is the most frequently mentioned theme across most models. 
\llama emphasized the most on \emph{removing unnecessary code}, specifically removing unnecessary or redundant variables or calculations. This suggests its tendency in cleaning up the code base in the hope of improving the energy efficiency of code. 
Mixtral placed a strong emphasis on memory optimization, \ie clearing unused memory and pre-allocating memory.
In addition to cleaning the code base, we found that the models focused on performance improvement, indicated by the theme \emph{improved code efficiency}.
Specifically, the LLMs focused on replacing inefficient functions with enhanced counterparts, such as built-in functions, more efficient commands, or better handling of data processing.
As Matlab is optimized for vectorized and matrix operations, we observed that \gpt-4 and \llama focused more on leveraging Matlab's optimized array and matrix operations reflected in themes such as \emph{vectorization} and \emph{improved indexing \& loops}.
In addition to code, \gpt-3 and \llama focused on improving code comments, specifically removing unnecessary or duplicate comments.
There were also instances where the LLMs did not find any optimization, \eg \llama indicated this 50 times.

\roundedbox{
\gpt-3, \gpt-4, and \llama suggested high-level code optimizations, while Mixtral provided code-specific suggestions targeting individual variables, functions \etc
}

The developer first independently established guidelines for energy optimization and applied those to 53 script.
\autoref{fig:themes-human-optimized-Files} shows the theme frequency in these script optimized by the developer. 
The distribution of themes in these script closely matches that of the LLM-optimized script in \autoref{fig:themes-all-models}.
Since every theme created from the guidelines is inherently inspected by the developer, no theme falls under \emph{no manual inspection} in this case.
We observed that the developer often did not focus on 
improving plots, matrix operations, error handling, or parallel processing.
When asked for reasoning, the developer explained that \emph{``parallel processing lowers execution time but not energy consumption''} and \emph{``error handling is rarely present in the scripts, so improvement in it might be insignificant''}.
Additionally, the developer mentioned a lack of knowledge in improving plots, specifically in optimizing the built-in plot functionality of Matlab.
Previous studies have highlighted similar reasons developers give regarding code sustainability \cite{Pang2016,rani2024energy}.
In some cases, the developer incorporated matrix, loop, and memory operations into broader themes like \emph{improved code efficiency}, \emph{vectorization}, and \emph{improved indexing \& loops}, so they were not classified independently.


Regarding the \emph{no optimization} category, we noted that \gpt-4 reported no cases where optimization is unnecessary or not possible, indicating its eagerness to always suggest solutions or even extend the code when it considers the current code incomplete.
In contrast, other models and the developer reported script where optimizations were unnecessary.
The developer used the \emph{improving code efficiency} theme twice as often as the LLMs, likely due to consistently replacing the function \texttt{xlsread} with \texttt{readmatrix}, as recommended by Matlab's documentation. LLMs appeared unaware of this official recommendation.

We observed that LLMs focused more often on improving code comments (theme \emph{improved comments}) while the developer neglected it often. For instance, LLMs translated code comments and variable names from Chinese to English. However, the Matlab developer avoided it to keep the original owner's thoughts intact. The developer only removed comments when they obstructed the readability aspect.
%
This observation underscores the differing priorities between automated and manual approaches. LLMs tend to generalize suggestions across languages or environments, potentially leading to overzealous changes, such as \gpt-4 always suggested optimizations or \llama translated code comments and variable names. Such changes reflect a focus on maintainability. In contrast, the developer's contextual understanding allowed him to focus on modifying only parts of the code they deemed necessary.

\roundedbox{
 LLMs often translated code comments and variable names, \eg Chinese to English. reflecting again a focus on readability and maintainability rather than energy efficiency. In contrast, the developer avoided translation to preserve the original author's intent,  removing comments only when they hindered readability.
}

\subsection{\texorpdfstring{$RQ_2$}{RQ2}: \rqIIIheader}
\label{subsec:rqIII-results}

In this RQ, we explored our hypothesis if the optimization applied by LLMs or humans enhance code's sustainability (\ie reduce energy, memory, and time consumption while ensuring that the optimized code is correct and still runs).
First, we check whether the optimized scripts not only affect the resource consumption, but preserve the correct behavior. 
\autoref{tab:Wilcoxon-energy-memory} shows the results for output matching ratio of original and optimized scripts (higher the better): the first row of the correctness consideration shows the `raw' correctness ratios, still containing all scripts. In the next row, we only compare the deterministic scripts' behavior; this is why the original non-deterministic scripts are set to score 1.00.
Overall, we find that one half to two thirds of the scripts still show the same behavior after optimization, with \gpt-3 and \gpt-4 yielding the most behavior-preserving optimizations.

\roundedbox{
\gpt-3 and \gpt-4 apply the most behavior-preserving optimizations, however, overall none of the approaches fully preserve the original behavior in one-third to half of the scripts.}

\begin{figure*}[tbh]
    \centering
        \includegraphics[width=\textwidth]{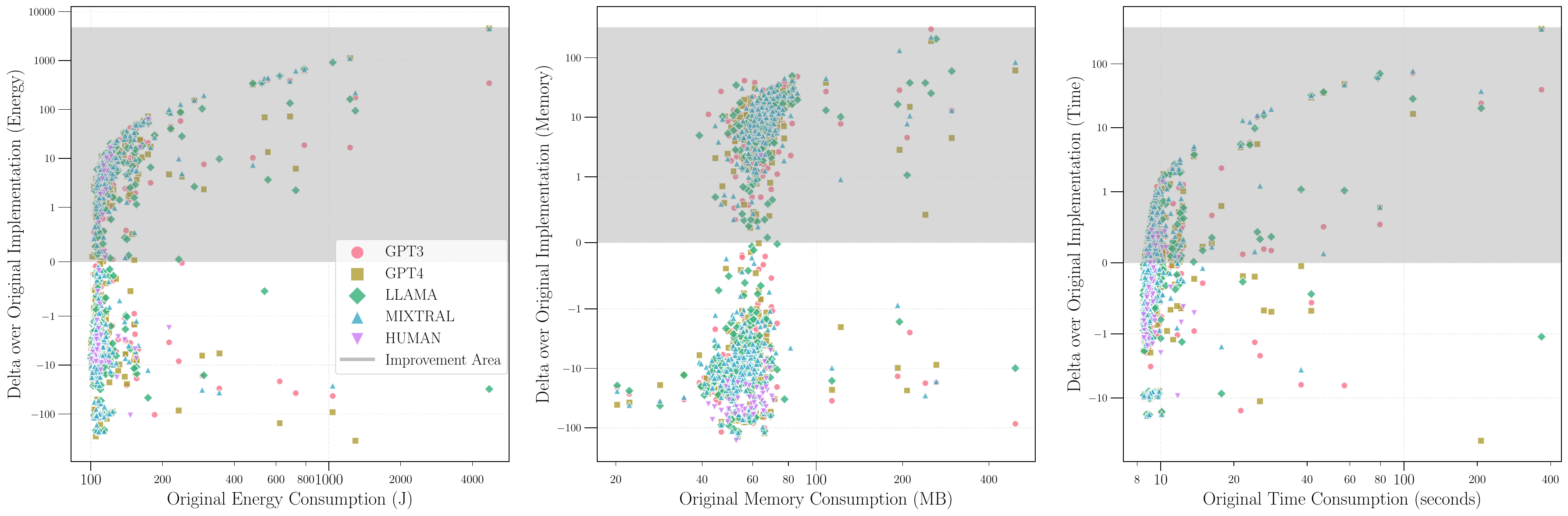}
        \caption{The plots show the delta (original - optimized implementation) achieved by LLMs and an senior developer. A positive delta (gray area) indicates improvement, \ie the optimized implementation is better. The y-axis is on a symmetrical logarithmic scale, and the x-axis on a logarithmic scale.}
        \label{fig:energy-memory-time-delta}
\end{figure*}

 The three sub-figures in \autoref{fig:energy-memory-time-delta} show the performance difference between the original implementation and the optimized versions (\ie original - optimized) generated by various models, \ie \gpt-3, \gpt-4, \llama, Mixtral and the developer, measured across the three main metrics energy consumption (left), memory consumption (middle), and execution time (right). 
In the figure, positive delta (in gray area) indicates that the optimized implementation outperformed the original by consuming less energy, time, or memory. 
The developer optimized a convenience sample of 53 script, while the models optimized at least 359 scripts. The data include Matlab's start-up and shutdown times, with the baseline overhead averaging 99.38~J, 51.07~MB, and 8.55~seconds. 

We calculated the central tendency (median) for various metrics, comparing the optimized script to the original ones.
The analysis showed that the median of the optimized versions tends to be higher than the original, indicating that optimization strategies often increase energy, memory, and time consumption, thereby actually reducing code sustainability. 
To further validate this finding, we performed a one-way Wilcoxon-signed rank test and calculated the effect sizes \cite{rosenthal1994parametric}.
Based on the results shown in \autoref{tab:Wilcoxon-energy-memory}, 
we can't conclude that any model consistently improved or worsened the code sustainability.

For energy consumption, none of the models showed a significant difference between the original and optimized versions (all $p$-values > 0.05).
We saw mixed effects: Mixtral and \llama showed positive effect sizes (0.025 and 0.079, respectively), suggesting lower energy consumption for the optimized versions, though not statistically significant.
In contrast, \gpt-3 and \gpt-4 showed negative effect sizes (-0.008 and -0.062, respectively), suggesting
higher energy consumption. 
Previous work showed that ChatGPT significantly outperformed other models in terms of sustainability consciousness~\cite{vartziotis2024learn}, however, our results contradict these.

For memory consumption, all models consistently showed increased usage in the optimized versions (all $p$-values < 0.05), with \llama showing a significant increase (effect size of -0.183).
Despite the fact that the models were instructed to focus on energy energy efficiency, many optimizations targeted memory, such as pre-allocating memory to variables or removing unnecessary code, but these efforts were ineffective or counterproductive. 
\roundedbox{
The energy optimizations showed no clear benefits to energy consumption, and had an unintended negative impact on memory consumption.
}
Regarding time consumption, we saw that \gpt-4 shows a statistically significant increase in execution time of the optimized code compared to the original one. 
Other AI models (\gpt-3, Llama, Mixtral) do not show statistically significant differences in execution time compared to the original code.
These findings suggest that the optimization strategies, whether automated or manual, might not fully account for the complexities of the Matlab environment, computing demands, and trade-offs.

\roundedbox{
\gpt-4 optimizations show a tendency to  
increase execution time, memory usage, and energy consumption. We observed a similar trend with developer optimizations.
}

Regarding the human-focused optimization, it had the largest effect (energy: -0.28, memory: -0.87, and time: -0.64) with a narrow confidence interval and statistical significance in all cases, indicating a substantial decrease in code's sustainability. 
Please note that this can be partly due to the small sample size (53 script), or that even senior developers might struggle to achieve energy saving without considering trade-offs (maintainability vs. energy efficiency).


\input{Tables/Wilcoxon-test-results}

We further explored the relationships among the metrics by calculating the Spearman correlation coefficients.
We observed a strong correlation between time and energy consumption across all models, indicating that longer execution times are closely linked to higher energy usage, emphasizing the need for optimization strategies that address both metrics simultaneously or provide more support in detangling this relation \cite{roque2024unveiling}.
The correlation coefficients are notably high and statistically significant.  For example, the correlation between the energy and time consumption in original scripts is $r = 0.77$, with \gpt-3 at $r = 0.82$, \gpt-4 at $r = 0.80$, \llama at $r = 0.84$, and Mixtral at $r = 0.77$. 
Moreover, the optimized energy consumption of \gpt-3 is highly correlated with that of \gpt-4 ($r = 0.58$), indicating similar energy consumption patterns between these models. 
Additionally, we observed a negative correlation between memory and time consumption of the original scripts ($r = -0.10$), suggesting that scripts with higher original memory usage have lower execution times or vice versa.
 These findings highlight the inter-dependencies between different performance metrics, providing insights into potential areas for further optimization.

\paragraph{Qualitative analysis of themes from LLMs optimization}
Our results show a strong correlation between time and energy; though this is not always consistent. Prior studies have demonstrated that a more time-efficient approach does not always result in the most energy-efficient solution \cite{pereira2017energy,pinto2014understanding,roque2024unveiling,strubell2020energy}. 
 Roque \etal further showed a non-linear relationship between energy and execution time. 
\cite{roque2024unveiling}.
To explore this relationship further, we conducted a qualitative analysis of script where the optimizations had conflicting effects on energy and execution time--either increasing energy consumption without reducing execution time or vice versa.

In RQ1, we identified various themes from the applied optimizations. Since multiple themes often influenced the energy efficiency of a script, isolating the impact of each theme was non-trivial. 
We specifically analyzed the distribution of PD (performance delta), which measure the differences between original and optimized versions. A positive PD indicates an improvement while a negative suggests the optimizations were ineffective.
Our focus was on the upper quartile (top 25\% of cases with the most improvement) and lower quartile (bottom 25\% of cases where performance worsened).
The top quartile themes included optimization like preallocation and vectorization across all metrics. In contrast, the bottom quartile featured themes such as redundant operations, complex logic, or failure to apply optimizations such as preallocating and indexing. 
\autoref{tab:performance-optimization-themes} shows these themes. 

\input{Tables/Performance-optimization-themes.tex}


In the trade-off analysis of energy and time, we found that common optimization themes included vectorization, preallocation, function replacement, and code reorganization. In the first case (optimizations saved energy but increased execution time), these themes reduced energy consumption by minimizing redundant computations and memory reallocation as seen in vectorized operations and preallocating arrays. However, it is possible that intermediate computation and overhead led to slower execution. Similarly, in the other case (optimizations increased energy but reduced execution time) these themes focused on improving execution time by eliminating loops or preallocating arrays but might have increased energy consumption due to memory cleanup. 

We also observed cases where the execution time seemed to be improving but had a negative impact on memory usage. For instance, reorganizing code, specifically moving clear statements to a later part and combining variable assignments seem to improve execution time in
one script
but had a negative impact on energy and memory, and in some cases, energy is impacted a lot. In script \texttt{unlinear\_regression.m}, LLM refactored code into a function. It may be the case that the function is further optimized by Matlab's JIT compiler that may have increased the energy consumption. 
These cases highlight the trade-offs among various sustainability metrics, energy, time, memory, or task accuracy -- in this case correctness.


\input{Tables/classification-human-optimized-files}

\paragraph{Qualitative analysis of themes from human optimization}
Similar to the analysis for LLMs in \autoref{tab:performance-optimization-themes}, we analyzed themes that surfaced in the human-optimized set of script (53 script) where energy and execution time showed opposite trends after optimization, \ie either energy consumption decreased while execution time increased (energy-optimized script), or vice versa (execution time-optimized script).
%
Table \ref{tab:classification-human-optimized-files} gives an overview of these 53 script.

In the correlation analysis (see \autoref{subsec:rqIII-results}), we observed a strong correlation between time and energy consumption across all models. In the human-optimized script,  37 out of 53 script (or 69.8\%) showed energy consumption and execution time changing in the same direction after optimization -- further confirming their strong correlation.
With the exception of one human-optimized script (\texttt{emplanner\_init.m}), all other 52 human-optimized script show a negative PD for memory, \ie they have used more memory after optimization. In the one script with positive PD, seven unused variables totaling 2MB were loaded. Removing their corresponding load statements likely contributed in reducing the memory consumption.

We analyzed the themes of the original and optimized script. We found that
the theme \emph{Improved Code Efficiency} appeared in five of the eight execution time-optimized script and in all three energy-optimized script. 
Analyzing the code changes we observed that preallocation might have an energy-optimized impact since dynamic reallocating memory can be avoided. Suppressing output to the command window is also likely to optimize the energy consumption.
In three out of five occurrences of this theme, an optimization involved the \texttt{load()} function.
The \texttt{load()}-function operates in two modes (see \autoref{lst:loadVariables}). In the first mode, all data is loaded into the current workspace, giving the developer no control over which variables are loaded into the workspace. In the mode 2.a), data can be assigned as struct fields to a specific variable. With the sub-mode 2.b) the developer can decide which fields should get loaded, allowing precise selection of the required variables from the file.
Although these two modes exhibit different behavior, our experiments did not provide conclusive evidence on whether one consumes less energy than the other. Future work can investigate their energy impact through specific controlled experiments.



The theme \emph{Removed Unnecessary Code} appeared in five energy-optimized script and in three execution time-optimized script. We expected this theme to reduce both energy consumption and execution time as executing fewer code fragments generally triggers fewer operations on the machine -- unless some special operation such as removing a \texttt{break} statement inside a for-loop could lead to more code being executed and increased execution time.
%
However, we could not always observe a reduction of both energy consumption and execution time. It might be that energy or time reduction got compensated by optimizations in connection with other themes.
In all cases where this theme appeared in energy-optimized script (where energy consumption decreased but execution time increased), the theme \emph{Improved Code Readability \& Maintainability} is also present, which we suspect contributes to increased execution time. Conversely, in the three execution time-optimized script (where energy consumption increased while execution time decreased), other code fragments likely counteracted the energy-saving effect of \emph{Removed Unnecessary Code} theme.
Specifically, two cases involved introducing temporary helper variables, and in one case, a call to \texttt{rng('default')} was introduced to set the random number generator seed during optimization.


The theme \emph{Improved Code Readability \& Maintainability} tends to lead to higher energy and execution time consumption, aligning with Cruz \etal's findings \cite{Luis19a}. 
For example, introducing helper variables improves maintainability but 
increases memory allocation and potentially increasing the energy consumption. This phenomenon could be observed, for example, in the script \texttt{MAIN\_minThrottleTraj.m}.
Similarly, extracting helper functions may improve maintainability but likely introduces execution time overhead.
%
However, we identified one case where improving maintainability likely reduced the execution time. The deprecated \texttt{ezplot} function shown in \autoref{lst:ezplot}, which takes a function description as a character vector and a definition interval for the plot as inputs, can be replaced with the recommended \texttt{fplot} function, using anonymous functions (see \autoref{lst:ezplot}).
Since \texttt{fplot} directly defines the function in code, it avoids the overhead of interpreting a character array 
by the Matlab-interpreter, likely reducing execution time.
We conducted a small energy consumption experiment to investigate these two functions further and it supports that \texttt{fplot} consumes up to 25\% less energy and up to 28\% less execution time than \texttt{ezplot}.

\smallskip
\begin{table}[h!]
    \centering
    \begin{minipage}{0.45\textwidth}
        \input{Listings/ezplotVsFplot}
    \end{minipage}
    \hspace{0.60cm}
    \begin{minipage}{0.45\textwidth}
        \input{Listings/loadVariables}
    \end{minipage}
\end{table}

The theme \emph{Improved Indexing \& Loops} is likely to reduce energy consumption. The code in \autoref{lst:loopWithSwitch} (from \texttt{BPDLX.m}) exhibits two inefficiencies: a switch-block inside a for-loop and absence of logical indexing. These issues are addressed  in the optimized version in \autoref{lst:loopWithoutSwitch}. An additional energy consumption experiment showed that the optimized code in \autoref{lst:loopWithoutSwitch} consumes only 22.1\% of the energy compared to the original code in \autoref{lst:loopWithSwitch}.
This highlights the energy saving potential of improved indexing and loop structures.
The theme \emph{Others} doesn't seem to have a consistent effect on either reducing or increasing energy consumption or execution time.

\begin{table}[h!]
    \centering
    \begin{minipage}{0.4\textwidth}
        \input{Listings/loopWithSwitch}
    \end{minipage}
    \hspace{0.60cm}
    \begin{minipage}{0.4\textwidth}
        \input{Listings/loopWithoutSwitch}
    \end{minipage}
\end{table}

\paragraph{Qualitative analysis of energy practices}
Next, we analyzed energy-intensive scripts with regard to optimization. We investigated five energy anti-patterns that we discovered in the scripts. The developer advised on how to improve the code fragments based on his experience and various Matlab guidelines.

\begin{itemize}
  \item \textit{Remove unnecessary code from loop:}
Code inside a loop that doesn't depend on the loop iteration should be moved outside to improve speed and reduce energy consumption (\eg \texttt{y = sin(x)} in \autoref{lst:redundant-code-inside-loop}). Although Matlab's code analyzer sometimes catches this, it misses most cases.

\item \textit{Remove redundant function evaluation:}
Evaluating a function \texttt{f(x)} both in an if-statement's condition and later in the if-statement's body is inefficient, leading to duplicate evaluations. 
A better practice is to assign the function results to a variable, \eg \texttt{helpVar = f(x)} outside the if-block and use this variable in both places. This approach reduces redundant function calls, particularly in loops, and enhances energy efficiency.

\item \textit{Replace inefficient calculation:}
Coding guidelines often advise avoiding unnecessary calculation\cite{higham2016matlab,matlab2024}. The following example is taken from the ``Algorithms\_MathModels'' project's \texttt{main.m} function \cite{reppackage}.
For instance, using $\texttt{tmp = randperm(n); x = tmp(1)};$ to generate a random integer $x$ between 1 and $n$ is inefficient if only \texttt{tmp(1)} is used, as the other $n-1$ numbers are unused. A more efficient approach is to use \texttt{x = randi(n)} directly.

\item \textit{Optimize set membership:}
Another recommendation is to carefully use \texttt{ismember} (to test for set membership). The following example is taken from the ``Algorithms\_MathModels'' project's \texttt{main.m} function.
This function is called roughly 300,000 times in nested loops, and thus makes the whole script inefficient. It is combinatorially expensive, especially when applied to a whole array to convert integers to a logical array as shown in \autoref{lst:ismember}. A more efficient approach is to use the variable \texttt{subset} directly as an index to truncate \texttt{selectedNumbers} (see \autoref{lst:truncateArray}). In an additional energy experiment we could show that the improved version from \autoref{lst:truncateArray} reduces the execution time by 22\%.

\item \textit{Use substractive over additive approach:}
Similarly, \autoref{lst:ismember} demonstrates an ``additive'' approach, where desired elements are added to an array, while \autoref{lst:truncateArray} uses a ``substractive'' approach, removing undesired elements. The latter method is more effective than the former in this case as it leverages the fact that the undesired elements are known in advance.
\end{itemize}

\begin{table}[h!]
    \centering
    \begin{minipage}{0.23\textwidth}
        \input{Listings/redundant-code-inside-loop}
    \end{minipage}
    \hspace{0.60cm}
    \begin{minipage}{0.35\textwidth}
        \input{Listings/ismember}
    \end{minipage}
    \hspace{0.15cm}
    \begin{minipage}{0.3\textwidth}
        \input{Listings/truncateArray}
    \end{minipage}
\end{table}

\section{Discussion and Implications}

\paragraph{Energy optimization guidelines}
In RQ$_1$, we collected various optimization guidelines that showed how the senior developer focused on narrow impactful optimizations, ignoring plotting or error handling, while LLMs focused on diverse optimization, ranging from readability to plotting to error management. This is partly due to limited knowledge about efficiency aspects~\cite{rani2024energy,pang2015programmers}. 
The developer acknowledged this gap and proposed hypotheses such as \emph{
    ``parallel processing reduce execution time so it may reduce energy consumption as well''}, which need further testing.
As Florath showed the potential of leveraging hybrid approaches (human-in-loop optimizations with LLMs) \cite{florath2023llm}, combining expert suggestions with LLM-generated optimization can overall improve code's energy efficiency.
%
Given that experts (or non-software engineers) in these domains often do not follow SE best practices~\cite{carver2007software}, they may rely more on LLMs for energy efficiency. 
This highlights the need for targeted guidelines and raising greater awareness among developers, and tailoring LLMs to follow these guidelines can improve recommendations and avoid unsupported suggestions.

\paragraph{Improving energy optimization via LLMs}
In our work, we tested both commercial and open source models. We observed that none of the tested LLMs were able to optimize energy consumption significantly. Even the presumably strongest model, \gpt-4 showed a negative effect, \ie performed worse than expected on energy and time consumption while Mixtral and \llama showed a positive effect regarding energy consumption. Florath \cite{florath2023llm} showed that Chat\gpt-4 improved performance in terms of time efficiency when a human-in-loop process is considered. We did not test this approach as our study scope was limited to assessing the effectiveness of models and comparing it to a typical developer approach.
In contrast, we found no significant time efficiency improvements from any model. Vartziotis \etal evaluated the sustainability of LLM-generated code and found the models lacking in this aspect~\cite{vartziotis2024learn}.
Our study also shows the same for Matlab. Our results show that the developer optimization did not reduce energy consumption, likely due to the developer’s limited knowledge of energy optimization and the absence of clear guidelines. Future work can explore establishing guidelines for energy efficiency.
Additionally, since only overall script optimizations were measured, individual optimizations may have canceled each other out.
Guidelines should account for the potential trade-offs between energy, time, and memory, as indicated by our results, to avoid worsening the code overall sustainability.
Our qualitative analysis found several themes and anti-patterns that may contribute to energy inefficiencies. 
Further experiments are required to verify these findings across other metrics, LLMs and programming languages.
We expect that fine-tuning these models with energy optimization patterns may help increase their energy awareness.

\paragraph{Evaluation of other languages and environments}
Despite Matlab's specialized toolboxes offering strong integration and performance optimization, the interest in other scientific computing platforms is growing. 
Python's open-source nature and libraries like Numpy and Pandas make it a popular alternative.
Similarly, R and Julia are emerging platforms for data visualization and high-speed computations respectively. 
However, even though some general rankings have been made~\cite{PEREIRA2021}, the energy efficient coding practices in these languages remains underexplored. Future work can extend our work to these languages.

\paragraph{Evaluation of entire workflows} 
Current tools focus on optimizing execution time and performance profiling, but still lack the means for measuring energy efficiency across entire workflows, from software requirement collection to deployment pipeline, due to the involvement of various components and unreliable metrics.
%
Also having the tools integrated within the Matlab platform,
to help Matlab developers streamline the process of identifying and optimizing energy inefficiencies. Such a tool could promote the adoption of energy-efficient practices and reduce the environmental impact of computing~\cite{grealey2022carbon}.

In our analysis of the top 100 projects, we observed diverse project characteristics, such as different domains (\eg scientific, engineering, utility), development activity cycles and pull requests, which may influence developer coding practices. Future work can explore whether green coding practices vary based on these characteristics and how to help non-software engineers accordingly to adopt these practices.
%

\section{Related Work}
With rising carbon emissions due to the technology sector, measuring and reducing software energy footprint is vital. 
Yet, there is a lack of guidelines and awareness among developers for making energy-efficient software~\cite{manotas2016empirical}.
Previous studies surveyed developers for software sustainability and found they rarely consider energy efficiency during development \cite{Pang2016,Shanbhag2022,manotas2016empirical} and lack awareness~\cite{rani2024energy} and tools to measure and develop energy-efficient software~\cite{manotas2016empirical}.
In our study, an expert Matlab developer was instructed first to establish the energy optimization guidelines and apply those to write energy-efficient code. The developer also reported the lack of official guidelines and knowledge in this regard.

As far as the energy impact of systems developed by non-software engineers is concerned, hardware aspects of specific domains, such as embedded systems, weather forecasting, physics, machine learning models, or computational models are mostly investigated~\cite{lacoste2019quantifying,lannelongue2021green,jumper2021highly,grealey2022carbon}. 

SE Researchers have measured energy consumption of software in various contexts, such as code execution \cite{ribic2014energy,pinto2014understanding}, third-party libraries \cite{schuler2020characterizing}, dataframe processing libraries \cite{shanbhag2022energy}, programming languages \cite{pereira2017energy}, Java applications \cite{vijaykrishnan2001energy}, embedded software \cite{tiwari1994power}, mobile applications \cite{kwon2013reducing}, or web applications~\cite{rani2024energy}.
As a result, many of them established energy patterns or anti-patterns~\cite{cruz2019catalog,Shanbhag2022,rani2024energy}, but none of them have focused on Matlab projects.
In our work, we measured the energy impact of Matlab code and identified related energy patterns, such as removing inefficient calculations or duplicate function evaluation.

Regarding the impact of using tools on code sustainability, 
Mehra \etal \cite{mehra2023assessing} identified energy-inefficient code using the CAST AIP tool and measured the impact of refactoring this code on software sustainability. 
With recent advancements in LLMs, they are also leveraged as tools for various SE tasks, including refactoring code and optimizing code~\cite{chen2024supersonic,sadik2023analysis,hou2023large,florath2023llm,vartziotis2024learn}.
Florath specifically explored the capabilities of using Chat\gpt-4 in an interactive workflow for optimizing computing efficiency for Python code~\cite{florath2023llm}. 
However, they focused on execution time efficiency and neglected energy efficiency.
Vartziotis \etal explored the capabilities of AI models, such as GitHub Copilot and Chat\gpt-3~\cite{vartziotis2024learn} in generating sustainable code. They found that LLMs lack sustainability awareness and are unable to reduce carbon emissions.
However, they did not analyze the types of optimizations applied by the models and neither assessed how closely these align with human optimizations.
Similarly, Cursaru \etal investigated the impact of source code generated by \llama \cite{cursaru2024controlled}.
In comparison, we measured the capabilities of various open-source and closed-source LLMs, \ie \gpt-3, \gpt-4, \llama, and Mixtral, for energy efficiency of Matlab code.
Furthermore, we compared the optimizations suggested by LLMs vs human and if they indeed impact the code sustainability of Matlab code.

Studies have explored the relation between time and energy complexity \cite{carter2023energy,brownlee2021exploring,carter2023energy} confirming a strong correlation. Our analysis yielded similar results, further analyzing the cases (energy vs. time) where one metric improved but other worsened. We highlighted themes, such as vectorization, preallocation influencing these trade-offs and suggest further investigation into their impact. Future work could examine string arrays vs. cell arrays \cite{near2021analysis} and vectorized vs. non-vectorized implementations \cite{sarkar2023computationally}, assessing their effects on both execution time and energy consumption. Additionally, the role of vectorization and preallocation in optimizing Matlab performance \cite{Bister2007IncreasingTS} warrants deeper exploration into their energy trade-offs.


\section{Threats to Validity}
\textit{Construct validity:}
The main concern for the construct validity is related to the measurements and definitions used in the study. 
While establishing the automated pipeline to measure the energy consumption of Matlab scripts, we identified and addressed various challenges. 
\begin{itemize}
\item \emph{Dependencies on Toolboxes:}
 Like other software, Matlab projects depend on various toolboxes or libraries, but identifying these dependencies in advance is difficult since Matlab lacks a built-in command for this. These dependencies complicate energy measurements, making it difficult to distinguish between energy usage from external dependencies and core functionalities.
We addressed this by installing all the available Matlab toolboxes. 

\item \emph{Code Abstraction:} Matlab's high-level abstraction helps non-software engineers write code without delving into low-level operations \cite{Leroy2021}, but this obscures the 
identification of energy-hungry code.
Our RQs aim to identify
coding practices that may impact energy efficiency and help establish a baseline for future comparison.

\item \emph{Entry-point Identification:} 
Automatically executing a project for energy measurement 
requires identifying an entry point script, which is a complex and tedious task. 
In this study, we developed heuristics to automatically identify entry points~\cite{reppackage}. 

\end{itemize}
In the optimization sessions, it is possible for a developer to get biased by interacting with LLMs or vice versa. To avoid both getting influenced from each other, we conducted both optimization sessions independently.
The inherent biases in training LLMs can result in suggesting certain optimization guidelines which might not reflect Matlab energy efficiency needs.
To mitigate this, we checked Matlab's official guidelines and inquired with the MathWorks support team. However, a lack of guidelines in this context restricted us in verifying all LLMs or developer suggested guidelines. 
Another concern is the definition of code's efficiency metrics that can influence our measurement, however these metrics have been used in previous work~\cite{florath2023llm,vartziotis2024learn}.

\textit{Internal validity:}
As Matlab provide only tool to measure execution time~\cite{matlab2024},
we measured energy consumption on a Linux server using the external tool EnergiBridge \cite{sallou2023energibridge}. This can introduce a bias for other platforms and tools.
To reduce the impact of these choices and reduce the impact of conflicting dependencies, we prepared a Docker image and used the platform-independent EnergiBridge that was used successfully in prior work~\cite{roque2024unveiling}.
To mitigate external factors like server load or warm-up periods, we repeated the execution 30 times and included a 5-second cool-down period.
Instead of giving the optimization guidelines ourselves, we collected them from LLMs and the human developer. Therefore the results can be influenced by the LLMs' or the developer's knowledge as other LLMs or developers might produce different results. To control this, we experimented with four LLMs and asked the developer to gain knowledge from the official guidelines and literature in this area.
One author manually mapped the themes from LLMs and the developer to 13 high-level themes, which may introduce a subjective bias. However, we mitigated this by following the closed-card sorting method, and involving another author in the theme categorization process.

\textit{External validity:}
The selection of the language, environment, projects, or script might introduce bias in our investigation. 
We chose Matlab due to its popularity for developing various AI and data-driven system by non-SE developers and the availability of a senior developer to us.
We mined and filtered the top 100 \github projects that met our criteria.
From these projects, we extracted script that were independent of user input and other scripts, and that could be automatically executed, filtering 1,937 scripts to 400.
We ensured that the number of script is above the sample set size of 343 script needed to draw statistically significant conclusions.
Similarly, the small sample (53 script out of 400) for human optimization can also impact our results. Since this is a time- and effort-intensive task, we suggest future work to expand on this analysis.
It is possible that the actual functionality occurred in the other script or functions called by these scripts.
Additionally, since only overall script optimizations were measured, individual optimizations may have canceled each other out.
Future work can focus on including the dependent scripts and functions as well. 

\textit{Conclusion validity:}
The accuracy and reliability of EnergiBridge can impact the validity of our conclusions about energy, time, or memory consumption. The tool is based on RAPL, often used for recording such measurements.
The developer's expertise can also impact other conclusions, \eg energy guidelines, optimization, or patterns. Therefore, we asked the developer to explore guidelines and resources about energy efficiency. 
Our study shows statistically insignificant results for energy consumption.
This raises concerns about either the absence of an effect, or the existence of an effect that is too subtle to catch or the need of a bigger sample size. 
Statistical power indicates the likelihood of detecting an actual effect in experimental research. Although the required statistical power to detect medium and large effect sizes has been met, smaller effects could go unnoticed. Therefore, we suggest future work to expand the analysis with a bigger sample size.


\section{Conclusion} 
As the carbon footprint of software development grows and LLMs become more prevalent, we explored
their effectiveness in optimizing Matlab code for energy efficiency.
Our study 
identified 2,176 unique frequent optimization themes from LLMs and mapped them to 13 higher-level themes. 
LLMs offered a broader range of optimizations--beyond energy efficiency--ranging from readability to error handling. Similarly, the senior developer focused narrowly on readability and efficiency aspects but overlooked aspects like error handling or parallel processing.
We also observed that LLMs, specifically \gpt-4, were eager to suggest optimizations or extend code even when other models and the developer suggested no optimization.

Finally, we measured the impact of these optimizations on energy, time, and memory. We found that the optimization had an unintended negative effect on memory usage without any clear benefits to energy and time. Although, time and energy consumption correlate, we identified various themes that influenced their trade-offs. We also identified various energy anti-patterns that may have contributed to the high energy consumption of the script.
This underscores the need for deeper analysis of complex programming operations and sustainability trade-offs.
Overall, our study highlights the need to evaluate the impact of various coding practices and raise awareness among developers and LLMs (via training and fine-tuning) to build sustainable software.

\section{Data Availability}
The dataset, software (measurement pipeline, and code used to analyze dataset), and the resultant optimizations from LLMs and the senior developer are available in our replication package \cite{reppackage}.


%% file: Listings/Prompt-llm.tex
\begin{minipage}{0.95\linewidth}
    \begin{lstlisting}[caption={An example prompt to LLMs},
     label={lst:Prompt-llm}]
System: You are an expert Matlab developer, who specializes in receiving and analyzing user-provided Matlab source code. Your primary function is to meticulously review the code, identify potential areas for energy optimization, and directly implement specific optimizations.

User: Optimize the Matlab code for energy efficiency, refactor it, and return the refactored code. Output first Matlab code starting with "```matlab" and ending with "```" and then the reasoning for the optimization.

Code: "{code_to_optimized}"
\end{lstlisting}
\end{minipage}

%% file: Listings/optimized-code-GPT-3.5.tex
\begin{minipage}{0.99\linewidth}
    \begin{lstlisting}[caption={\gpt-3 Response for \autoref{lst:Prompt-llm}},
     label={lst:optiimized-PSO-function-gpt3}]
```matlab
{Optimized_code}
```
Reasoning for the optimization:
1. Parameterization: By storing the varying parameters in arrays, we can reduce code duplication and make it easier to modify the parameters in the future.
2. Loop Optimization: Instead of repeating the same code with different parameters multiple times, we can use a loop to iterate over the parameters and optimize the code execution.

\end{lstlisting}
\end{minipage}


%% file: Tables/Themes-frequency.tex
\begin{table}
\centering
\scriptsize
		\caption{Themes proposed by various LLMs for optimizing 400 files. We extracted unique themes from total themes and mapped them manually to high-level themes.}
  \label{tab:reasoning-themes-frequency}
\begin{tabular}{lrrrr}
    \toprule
     \textbf{Theme extraction steps}  & \textbf{\gpt-4} & \textbf{\gpt-3} & \textbf{\llama} & \textbf{Mixtral}  \\ 
     \midrule
    Optimized Files  & 386 & 387 & 398 & 393  \\ 
    Files with Reasoning  & 373 & 377 & 392 & 393 \\ 
    Total Themes  & 1,632 & 1,811 & 1,998 & 2,894  \\ 
    Unique Themes & 719 & 1,086 & 703 & 1,335 \\
    Manually Mapped Themes & 360 & 582 & 541 & 693 \\
    High-level Themes & 13 & 13 & 13 & 13 \\
    \bottomrule
\end{tabular}
\end{table}

%% file: Tables/Optimization-categories.tex
\begin{table*}
\scriptsize
\centering
		\caption{High-level energy optimization themes by LLMs and the developer with their description.}
  \label{tab:optimization-categories}
		\begin{tabular}{p{0.2\linewidth}p{0.75\linewidth}}			\toprule
     \textbf{Theme by LLMs} &  \textbf{Description with reasoning}  \\ 
     \midrule
Improved Code Readability \& Maintainability   & Improve code by organizing, refactoring, following consistent naming conventions, or formatting for better readability and maintainability \\
\midrule

Removed Unnecessary Code   & Remove unnecessary or redundant code, \eg variables, methods, calculations\\
\midrule

Improved Code Efficiency   & Replace functions or algorithms with their efficient counterparts (\eg Replacing `xlsread' by `readtable' for file reading), use built-in functions, optimize data processing to improve code's energy efficiency 
\\\midrule

Improved Memory Management   &  Improve memory efficiency by pre-allocating, pre-computing, or clearing variables to reduce memory usage and avoid dynamic resizing \\
\midrule

Vectorized Operations   & Use vectorized operations to improve performance as Matlab is optimized for vectorized computation \\
\midrule

Improved Indexing \& Loops   & Optimize loops and indexing of variables or arrays  \\
\midrule

Improved comments    & Remove unnecessary or redundant comments, add or update comments for readability and understanding \\
\midrule

Improved Plots    & Improve titles or legends of plots, remove or consolidate figures\\
\midrule

Improved Matrix Operations   & Make use of matrix-related operations, such as transposition or matrix multiplication  \\
\midrule

Improved Error Handling    & Improve exception handling, error messages and try-catch blocks to prevent potential errors and help finding their root cause \\
\midrule

Parallel processing    & Using \emph{parfor} instead of \emph{for} to enable parallel processing of independent iterations on multi-core machines.\\
\midrule


Others    & Insufficient context for mapping to existing high-level themes. \\
\midrule

No Optimization   & No changes to the code \\

\midrule

No Manual Inspection   & Less-frequent themes, often appearing only once and related to specific code, not manually inspected for mapping to the existing high-level themes. \\

\hline
\end{tabular}
\end{table*}

%% file: Tables/Wilcoxon-test-results.tex
\begin{table}[t]
\centering
\scriptsize
\caption{Results of statistical tests for evaluating code's green capacity.
Energy, Memory, and Time are tested using a one-sided Wilcoxon test (Hypothesis: optimization decreases code's green capacity) while correctness is tested by matching the outputs of both executed script versions. In the correctness analysis, the first row is before adjusting for non-determinism.}
	\label{tab:Wilcoxon-energy-memory}
\begin{tabular}{lc|c|c|c|c}
\toprule
Metric & GPT-3 & GPT-4 & Llama & Mixtral & Human \\
\midrule
& \multicolumn{5}{c}{(p-value, Effect Size)} \\
\cmidrule(lr){2-6}
Energy & 0.439, -0.008 & 0.119, -0.062 & 0.686, 0.025 & 0.934, 0.079 & 0.025, -0.278 \\
Memory & 0.006, -0.133 & 0.013, -0.117 & 0.000, -0.183 & 0.033, -0.096 & 0.000, -0.869 \\
Time & 0.165, -0.051 & 0.012, -0.119 & 0.237, -0.042 & 0.543, 0.006 & 0.000, -0.637 \\
\cmidrule(lr){2-6}
 & \multicolumn{5}{c}{(Output matching ratio)} \\
\cmidrule(lr){2-6}
Correctness (0.73) & 0.46   & 0.47   & 0.39   & 0.37    & 0.29\\
Non-determ. accounted (1.00) & 0.64 & 0.64 & 0.54 & 0.50 & 0.58\\

\bottomrule
\end{tabular}

\end{table}

%% file: Tables/Performance-optimization-themes.tex
\begin{table*}
\scriptsize
\centering
		\caption{Common optimization themes (by LLMs) based on performance delta (PD = original-optimized) for energy and time trade-off where \textcolor{green}{$\uparrow$} denotes the upper quartile (top 25\% improvement) and  \textcolor{red}{$\downarrow$} denotes the lower quartile (bottom 25\% decline)}
  \label{tab:performance-optimization-themes}
		\begin{tabular}{p{0.12\linewidth}p{0.23\linewidth}p{0.59\linewidth}}			\toprule
       \textbf{Case} & \textbf{Themes} &  \textbf{Description}  \\ 
     \midrule
 \textcolor{green}{$\uparrow$} Energy, \textcolor{red}{$\downarrow$} Time &  Vectorization & Replacing loops with vectorized operations as seen in \texttt{ft\_postamble\_hastoolbox.m}, \texttt{Evaluation\_for\_Multi\_Algorithm.m}\\
\cline{2-3}
 &Improved Memory Management & Preallocating arrays for batch processing or avoiding dynamic resizing, using \texttt{find} instead of loop\\
\cline{2-3}
& Improved Code Efficiency & Replacing custom loops and functions with built-in functions like \texttt{randi, randsample, fprintf} like in \texttt{elm\_stock.m}, Replacing \texttt{ismember} function with \texttt{setdiff}, Replacing individual assignments with \texttt{struct} initialization, Replaced, built-in functions \texttt{sum} with \texttt{mean}, Converting structure array to cell array\\
\cline{2-3}
& Removed Unnecessary Code & Removing redundant operation such as in \texttt{chap11\_8.m}  \\
\cline{2-3}
 &  Parallel Processing & switch from \texttt{for} to \texttt{parfor} improved energy but introduced synchronization overhead \\
\cline{2-3}
 & Improved Code Readability \& Maintainability & Adding and Removing comments, Translating comment from other languages (Chinese) to English, formatting, spacing \\
\midrule

    \textcolor{red}{$\downarrow$} Energy, \textcolor{green}{$\uparrow$} Time  

    &  Vectorization  & Using vectorized functions like \texttt{ndgrid} instead of \texttt{meshgrid}, replaced the explicit array definition with \texttt{linspace} in \texttt{emplanner\_init.m} \\
\cline{2-3}
 &  Improved Memory Management & Preallocating arrays (Used `linspace` to generate `vx\_break\_point` instead of a loop) like in \texttt{emplanner\_init.m, example\_quivers.m, chap6\_16.m},  \\
\cline{2-3}
 &  Improved Code Efficiency & Replaces code with in-place operations, Operators (`./` operator than the `/` operator), or Replacing built-in functions \texttt{sum} with \texttt{mean}, replacing `clear all` with `clear` \\
\cline{2-3}
 & Removed Unnecessary Code & Removing unnecessary variables such as in \texttt{example\_quivers.m}, Reduced function call  \\
   \cline{2-3} 
  & Improved Code Readability \& Maintainability & Simplified calculations, reorganized code or combined variables assignments, refactoring code into a function \eg in script \texttt{distortion\_saver.m, MAIN.m, chap15\_3.m}, Adding, or removing comments, or translating comment from other languages (Chinese) to English, formatting code, adding semicolons \\
\cline{2-3}
& Improved Matrix Operations & Replaced functions for matrix operations like in \texttt{Ergodic\_Capacity\_Correlation.m} \\
  

\hline
\end{tabular}
\end{table*}

%% file: Tables/Classification-human-optimized-files.tex
\begin{table}
\scriptsize
\centering
\caption{Classification of human-optimized files based on the performance delta (PD=original-optimized) for Energy and Time Trade-off where $\uparrow$ denotes PD is positive and $\downarrow$ denotes PD is negative.}

    \begin{tabular}
    {lr}
        \textbf{Case} &  \textbf{
        \# Files}  \\ 
        \midrule
        $\uparrow$ Energy, $\downarrow$ Time (energy-optimized files) & 8 \\
        $\downarrow$ Energy, $\uparrow$ Time (runtime-optimized files) & 3 \\
        $\uparrow$ Energy, $\uparrow$ Time (energy-time-optimized files) & 8 \\
        $\downarrow$ Energy, $\downarrow$ Time (non-optimized files) & 29 \\
        Unusable & 5 \\
        \textbf{Total} & \textbf{53} \\
        \bottomrule
    \end{tabular}\label{tab:classification-human-optimized-files}
\end{table}

%% file: Listings/ezplotVsFplot.tex
\begin{minipage}{\linewidth}
    \begin{lstlisting}[caption={Plotting functions ezplot() and fplot()},label={lst:ezplot}]
% deprecated function for plotting:
ezplot('x*sin(10*pi*x)+2',[-1,2])

% recommended function for plotting:
fplot(@(x)x .* sin(10 * pi * x) + 2,[-1,2])
    \end{lstlisting}
\end{minipage}

%% file: Listings/loadVariables.tex
\begin{minipage}{\linewidth}
    \begin{lstlisting}[caption={Different operating modes of the \texttt{load()}-function},label={lst:loadVariables}]
% mode 1) load data into workspace
load('filename.mat')

% mode 2.a) load data and assign to variable
var = load('filename.mat');

% mode 2.b) load data and assign specific
% field to variable
var = load('filename.mat').fieldName;
    \end{lstlisting}
\end{minipage}

%% file: Listings/loopWithSwitch.tex
\begin{minipage}{0.99\linewidth}
    \begin{lstlisting}[caption={Loop with switch},label={lst:loopWithSwitch}]
% set random generator seed
rng default
% generate array of indices in range 1,...,4
indices=randi(4,2000,1);

% fill array output
for i=1:2000
    switch indices(i)
        case 1
            output(i,:)=[1 0 0 0];
        case 2
            output(i,:)=[0 1 0 0];
        case 3
            output(i,:)=[0 0 1 0];
        case 4
            output(i,:)=[0 0 0 1];
    end
end
    \end{lstlisting}
\end{minipage}

%% file: Listings/loopWithoutSwitch.tex
\begin{minipage}{0.99\linewidth}
    \begin{lstlisting}[caption={Loop without switch},label={lst:loopWithoutSwitch}]
% set random generator seed
rng default
% generate array of indices in range 1,...,4
indices = randi(4,2000,1);

% fill array output
output = zeros(2000,4);
for idx1 = 1:4
    output(indices == idx1,idx1) = 1;
end
    \end{lstlisting}
\end{minipage}

%% file: Listings/redundant-code-inside-loop.tex
\begin{minipage}{0.99\linewidth}
\begin{lstlisting}[caption={Unneccesary code inside a for loop},label={lst:redundant-code-inside-loop}]
x = 1:0.01:2 * pi;
n = 100;
values = zeros(1,n);
for idx = 1:n
    y = sin(x);
    values(idx) = y(idx);
end
\end{lstlisting}
\end{minipage}

%% file: Listings/ismember.tex
\begin{minipage}{0.99\linewidth}
    \begin{lstlisting}[caption={Inappropriate use of \texttt{ismember}},label={lst:ismember}]
allNumbers = 1:31;
subset = [20,21,22];
index = ~ismember(allNumbers,subset);
selectedNumbers = allNumbers(index);
    \end{lstlisting}
\end{minipage}

%% file: Listings/truncateArray.tex
\begin{minipage}{0.99\linewidth}
    \begin{lstlisting}[caption={Better apply array truncation},label={lst:truncateArray}]
allNumbers = 1:31;
subset = [20,21,22];
selectedNumbers = allNumbers;
selectedNumbers(subset) = [];
    \end{lstlisting}
\end{minipage}